\pdfoutput=1

\documentclass[aps,prl,twocolumn,superscriptaddress]{revtex4-1}

\usepackage{amsmath} 
\usepackage{graphicx} 
\usepackage{color}
\usepackage{siunitx}
\usepackage[dvipsnames]{xcolor}
\usepackage[normalem]{ulem}

\definecolor{red}{rgb}{0,0,0}
\newcommand{\red}{\color{red}}
\definecolor{blue}{rgb}{0,1,0}

\mathchardef\mhyphen="2D

\begin{document}

\title{Superconductivity in the regime of attractive interactions in the Tomonaga-Luttinger liquid}

\author{\v Z. Gosar}
\affiliation{Institute Jo\v zef Stefan, 
Jamova c.~39, 1000 Ljubljana, Slovenia}
\affiliation{Faculty of mathematics and physics, University of Ljubljana, 
Jadranska c.~19, 1000 Ljubljana, Slovenia}
\author{N. Jan\v sa}
\affiliation{Institute Jo\v zef Stefan, 
Jamova c.~39, 1000 Ljubljana, Slovenia}
\author{T. Arh}
\affiliation{Institute Jo\v zef Stefan, 
Jamova c.~39, 1000 Ljubljana, Slovenia}
\author{P. Jegli\v c}
\affiliation{Institute Jo\v zef Stefan, 
Jamova c.~39, 1000 Ljubljana, Slovenia}
\author{M. Klanj\v sek}
\affiliation{Institute Jo\v zef Stefan, 
Jamova c.~39, 1000 Ljubljana, Slovenia}
\author{H. F. Zhai}
\affiliation{The University of Texas at Dallas,  800 West Campbell Road Richardson, Texas 75080-3021, USA}
\author{B. Lv}
\affiliation{The University of Texas at Dallas,  800 West Campbell Road Richardson, Texas 75080-3021, USA}
\author{D. Ar\v con}
\email{denis.arcon@ijs.si}
\affiliation{Institute Jo\v zef Stefan, 
Jamova c.~39, 1000 Ljubljana, Slovenia}
\affiliation{Faculty of mathematics and physics, University of Ljubljana, 
Jadranska c.~19, 1000 Ljubljana, Slovenia}

\begin{abstract}

  {
While the vast majority of known physical realizations of the Tomonaga-Luttinger liquid (TLL) have repulsive interactions defined with the dimensionless interaction parameter $K_{\rm c}<1$, we here report that  Rb$_2$Mo$_3$As$_3$ is in the opposite TLL regime of attractive interactions. This is concluded from a TLL-characteristic  power-law temperature dependence of the $^{87}$Rb spin-lattice relaxation rates over broad temperature range yielding the TLL interaction parameter for charge collective modes   $K_{\rm c}=1.4$.  The TLL of the one-dimensional band can be traced almost down to  $T_{\rm c} = 10.4 $~K, where the bulk superconducting state  is stabilized by the presence of a three-dimensional band and characterized by the $^{87}$Rb temperature independent Knight shift and the absence of Hebel-Slichter coherence peak in the relaxation rates. The  small superconducting gap measured in high magnetic fields reflects either the importance of the vortex core relaxation or  
the uniqueness of the superconducting state stemming  from the attractive interactions defining the precursor TLL.
}
\end{abstract}



\maketitle


A universal paradigm of the Tomonaga-Luttinger liquid (TLL)  \cite{giamarchi2003, Anderson1D,  schulz1998, Ohta2008, pouget2012, schonhammer2004}  describes the physics of interacting fermions  in one dimension remarkably well and predicts their most characteristic features that can be directly verified in experiments: collective excitations which generally separate into spin and charge modes and a power-law decay of the spin and charge correlation functions at long distances that lead to  power-law dependencies of the corresponding experimental quantities as a function of temperature or frequency.  Experimentally, such power-law dependencies were taken as a hallmark of TLL physics in molecular conductors such as tetrathiafulvalenetetracyanoquinodimethane TTF-TCNQ or Bechgaard salts (TMTSF)$_2$X, where TMTSF denotes tetramethyltetraselenafulvalene \cite{Jerome}, carbon nanotubes \cite{Bockrat, Ihara_2010} and in one-dimensional antiferromagnetic insulators \cite{ladder, CsO2_PRL, CsO2_PRB}. 

These physical realizations of the TLL may be very diverse, but their behavior is universal as the extracted power-law exponents are  functions of only the TLL parameter $K$, which characterizes the sign and the strength of  interactions between collective excitation modes  \cite{giamarchi2003}. While the interactions in the TLLs are generally repulsive with $K<1$, the only example of  attractive interactions with $K>1$ has been found in the quantum spin ladder system (C$_7$H$_{10}$N)$_2$CuBr$_4$ \cite{ladder, Jeong_PRL_2016}.  
In one-dimensional metals, $K$ depends on the strength of various scattering processes \cite{Solyom1979}, which also define the competing orders: charge density wave, spin density wave or superconductivity. Remarkably,  both  the singlet and the triplet  superconducting fluctuations are predicted \cite{giamarchi2003, Anderson1D,  schulz1998, Ohta2008, pouget2012} in the regime of attractive interactions. 
The emerging superconductivity in the $K>1$ regime thus remains experimentally  unexplored. 

Here we focus on an intriguing class of solids, which are derived from the self-assembly of chains with inherently weak interchain coupling. Molybdenum-chalcogenide (Li$_x$MoS$_2$, Na$_{2-\delta}$Mo$_6$Se$_6$)  and molybdenum-oxide (Li$_{0.9}$Mo$_6$O$_{17}$) chains \cite{LiMoS_PRl, NaMoSe_NC, LiMo_NC, LiMo_PRB} display many unusual properties associated with  TLLs, but they also show a high degree of disorder due to the intercalation of metallic atoms between the chains. 
The disorder effects seem to be less important in Cr-pnictide  quasi-one-dimensional metals, e.g., A$_2$Cr$_3$As$_3$ (A=K, Rb, Cs) with Cr$_3$As$_3$ chains as main building units, which remarkably also  show low-temperature superconductivity with the critical temperature  $T_{\rm c}\approx 5$~K \cite{Cao_PRX2015, Tang-RbCr, KCrPRL2017, Zhi_PRB,Yuan2015,  KCr_muSR_PRB, KCr_muSR, Imai_PRL, Mu-KCr3As3}. There seems to be a growing consensus for the unconventional singlet superconducting states in this family of materials based on the large specific-heat jump at $T_{\rm c}$ and large upper critical fields exceeding Pauli limit \cite{Cao_PRX2015, Tang_2017}, absence of the Hebel-Slichter coherence peak and the power-law dependence of nuclear spin-lattice relaxation rates, $1/T_1$, \cite{Imai_PRL, Zhi_PRB} and the penetration depth proportional to temperature \cite{Yuan2015} that implies the nodal-type of a superconducting gap. Moreover, a possible ferromagnetic quantum critical point has been proposed based on variation of $1/T_1$ with the Cr-As-Cr angle \cite{CrAs_FM_SC}.  However, the complicated electronic structure comprising two quasi-one-dimensional (1D) and one three-dimensional (3D) Fermi surface \cite{ElStr_Cr2As3} and the presence of a Korringa component in the $1/T_1$ cast some doubts on the direct applicability of the TLL model for A$_2$Cr$_3$As$_3$ \cite{Zhi_PRB}. 

\begin{figure} [tb]
  \centering
  \includegraphics[width=1.0\linewidth]{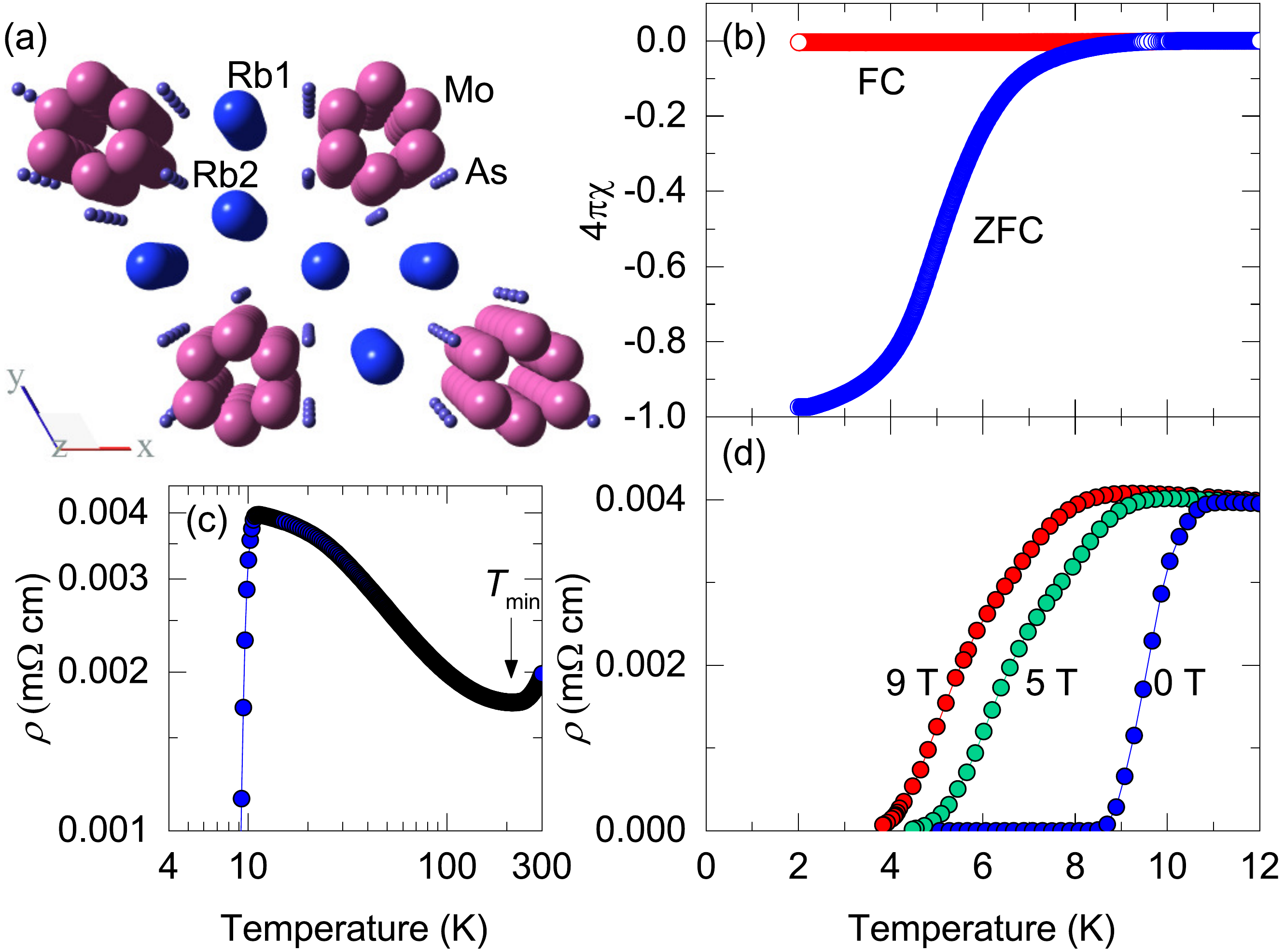}
  \caption{
(a) The assembly of  Mo$_3$As$_3$ chains in the Rb$_2$Mo$_3$As$_3$  structure. Large pink spheres stand for Mo atoms whereas small violet spheres for As atoms. Rb atoms (large blue spheres) 
are in the two different crystallographic positions Rb1 ($3k$) and Rb2 ($1c$).   (b) Temperature dependence of static spin susceptibility of Rb$_2$Mo$_3$As$_3$ measured under the zero-field-cooled (blue) and the field-cooled (red) protocols at $\mu_0H=1$~mT.  (c) Temperature dependence of zero-field resistivity  $\rho(T)$ (blue circles) showing a superconducting transition at   $T_{\rm c}(0)=10.4$~K. (d) Low-temperature resistivities measured in zero-field (blue), 5~T (green) and 9~T (red).} 
  \label{fig0}
\end{figure}

Recently, superconductivity has been discovered in K$_2$Mo$_3$As$_3$ made of assembled Mo$_3$As$_3$ chains (Fig.~\ref{fig0}a) with a relatively high zero-field  $T_{\rm c}(0)=10.4$~K   \cite{Mo3As3}. Although A$_2$Mo$_3$As$_3$ are isostructural to their Cr counterparts A$_2$Cr$_3$As$_3$ and the first principle calculations suggest that they share similar electronic structure \cite{Yang2019}, the $4d$  character of electrons and stronger spin-orbit coupling may separate this family of superconductors into its own class.   Here we report on the normal and superconducting state in  Rb$_2$Mo$_3$As$_3$   and find some fundamental differences with respect to the family of  A$_2$Cr$_3$As$_3$.  Strikingly, the TLL  of the 1D components of the Fermi surface in which effective attractive   interactions between  collective charge excitations prevail  can be followed in $^{87}$Rb nuclear magnetic resonance (NMR) experiments  over an extremely broad temperature range from $\sim 200$~K almost down to  $T_{\rm c}$.  On the other hand, $^{75}$As nuclei couple more strongly to the 3D band.  Directly from such a multi-orbital state, superconductivity develops, but the insensitivity of the  $^{87}$Rb Knight shift to the onset of the superconducting state, the absence of Hebel-Slichter coherence peak and the reduced superconducting gap  in a moderate field imply its unconventional character.

 Powder samples of Rb$_2$Mo$_3$As$_3$ were prepared via standard high-temperature solid state method previously used to synthesize A$_2$Cr$_3$As$_3$ samples \cite{Tang-RbCr, Mu-KCr3As3}. Powder X-ray diffraction measurements show phase pure sample with the A$_2$Cr$_3$As$_3$-type structure -- hexagonal crystal lattice with a space group of $P\overline{6}m2$ \cite{Cao_PRX2015}. The sample was  further investigated by dc magnetic susceptibility measurements in the zero-field-cooled (ZFC) and field-cooled (FC) modes at  $\mu_0H=1$~mT (Fig.~\ref{fig0}b). The Rb$_2$Mo$_3$As$_3$ compound indeed shows a bulk superconductivity below zero-field $T_{\rm c}(0) \approx 10$~K with the large diamagnetic shielding $4\pi\chi_{\rm ZFC}$ of 97\% at 2~K demonstrating sample's high quality. Next, the temperature dependence of the zero-field resistivity, $\rho(T)$, was measured (Fig.~\ref{fig0}c) using  Quantum Design Physical Property Measurement System (PPMS) system. The resisitivity at high temperatures  decreases with decreasing temperature, with a minimum at $T_{\rm min}\sim 210$~K where $\rho(T)$ starts to increase with decreasing temperature. Just above $T_{\rm c}(0)=10.4$~K, $\rho(T)$ first almost flattens as a function of temperature and then it sharply drops to zero where the superconducting state sets in. The  normal-state increase in $\rho(T)$ with decreasing temperature  below $T_{\rm min}$  may imply the weak localization effects due to the presence of disorder and Coulomb repulsion \cite{Anderson1D, NaMoSe_NC}. 
 Measurements of  $\rho(T)$  in the magnetic field show a suppression of superconductivity and the corresponding decrease of $T_{\rm c}$ with increasing magnetic field (Fig.~\ref{fig0}d). In the field of $9$~T, the critical temperature decreases to  $T_{\rm c}\approx 8$~K. The magnetization and resistivity data thus confirm the high quality of the sample and the emerging superconducting state, but could not resolve alone on  the potential TLL physics  in this material nor on the symmetry of the superconducting state. 
 
\begin{figure} [tb]
  \centering
\includegraphics[width=1.0\linewidth]{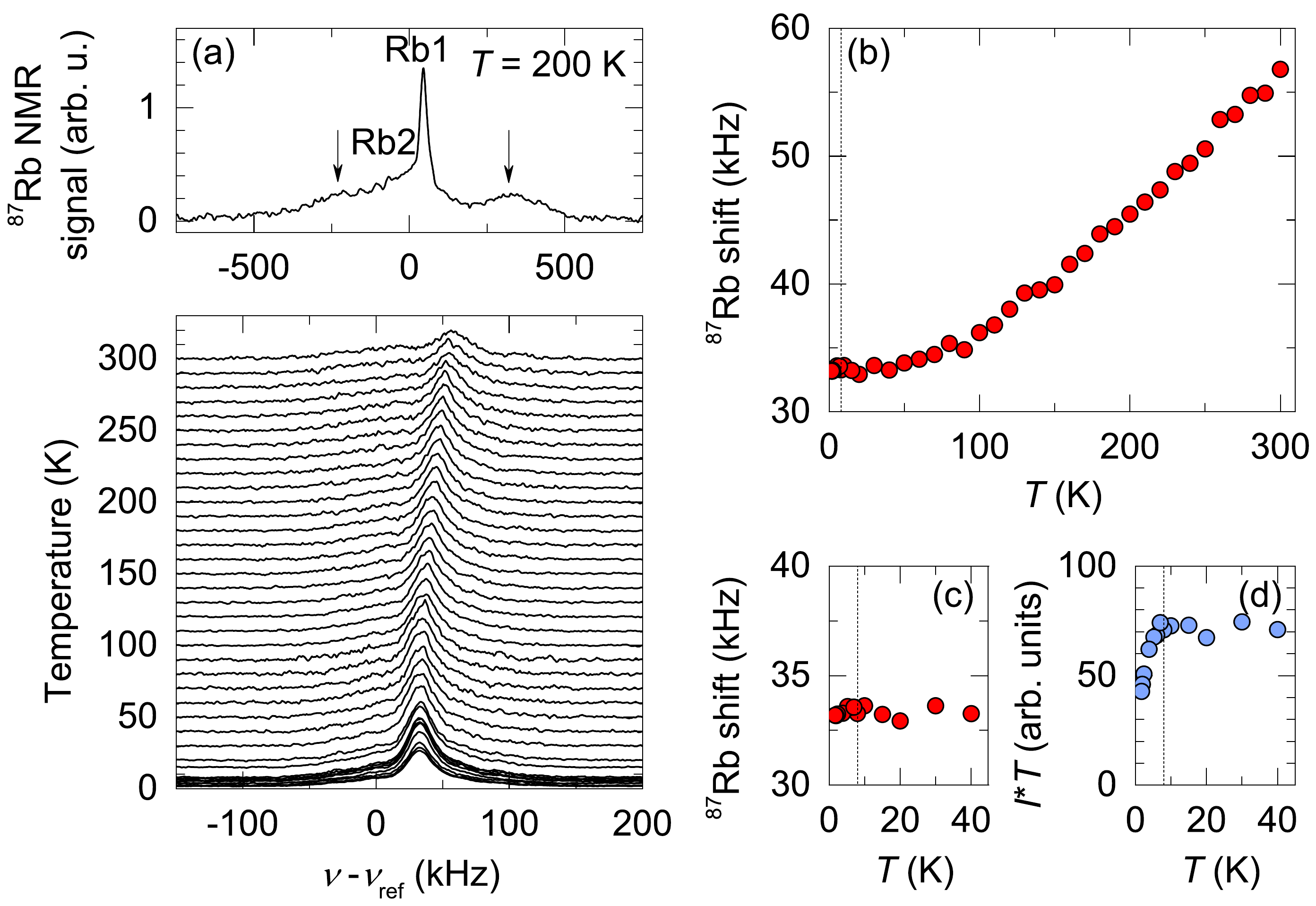}
  \caption{
  (a) $^{87}$Rb  NMR spectra in Rb$_2$Mo$_3$As$_3$ powder. 
  Top: the full spectrum measured at $T=200$~K. Arrows indicate the two satellite transitions at $\pm \nu_{\rm Q}$ relative to the Rb1-site central 
transition. The region of the non-shifted Rb2 component is also indicated. Bottom: the full temperature dependence of the Rb1 central transition line. (b) Temperature dependence of the $^{87}$Rb NMR  shift for Rb1 site. (c) Temperature dependencies  close to $T_{\rm c}$ for the  $^{87}$Rb NMR  shift  and  (d) for the integrated signal intensity multiplied by temperature.  $T_{\rm c}\approx 8$~K (dashed vertical lines)  in the NMR field of 9.39~T. 
  }
  \label{fig1}
\end{figure}

We  thus  turn to $^{87}$Rb (nuclear spin $I=3/2$) NMR, which was measured in magnetic fields of  4.7~T and 9.39~T with the corresponding reference Larmor frequencies of  $\nu_{\rm L}=65.442$~MHz and $130.871$~MHz, respectively. We note that Rb atoms are   intercalated into the voids between the Mo$_3$As$_3$ chains (Fig.~\ref{fig0}a) and 
 thus couple to the intrinsic  intra-chain low-energy dynamics through the contributions of the Mo $4d$ bands that cross the Fermi energy \cite{Yang2019}.
The $^{87}$Rb NMR spectrum measured in 9.39~T (Fig.~\ref{fig1}a)  is composed of a narrow peak corresponding to the central $1/2\leftrightarrow -1/2$ transition, which is  shifted for $\sim 56$~kHz with respect to the $\nu_{\rm L}$, and the two broadened satellite peaks that symmetrically flank the central one and which yield the quadrupole frequency $\nu_{\rm Q} \approx 280$~kHz. The second-order quadrupole shift of the central transition line, $\delta\nu\propto \nu_{\rm Q}^2/\nu_{\rm L}$, is thus small. The measured  line shift must therefore come almost entirely  from the  hyperfine coupling to the itinerant electrons, {\em i.e.}, the Knight shift $K_{87}$.
In addition to this $^{87}$Rb resonance, we find also a broader component with a  small shift and whose satellite transitions cannot be resolved  (Fig.~\ref{fig1}a). The two-component NMR spectrum reflects  two crystallographic  Rb sites (at the $3k$ and $1c$ positions \cite{Cao_PRX2015}, Fig.~\ref{fig0}a), where only one of them (probably  Rb1 at the $3k$ positions)  strongly couples to the  Mo$_3$As$_3$ chains. The analogous two-component $^{133}$Cs NMR spectrum has been observed also in  Cs$_2$Cr$_3$As$_3$ \cite{Zhi_PRB}.

 
The central transition Rb1 peak strongly shifts with temperature between room temperature and 
$T=50$~K and then becomes almost temperature independent between 50~K and $T_{\rm c}$. Since the Knight shift, $K_{87}={a_{\rm Rb}\over N_{\rm A} \mu_{\rm B}}\chi(T)$ (here  $a_{\rm Rb}$, $N_{\rm A}$ and $\mu_{\rm B}$ are the $^{87}$Rb hyperfine coupling constant, the Avogadro number and the Bohr magneton, respectively),  is directly proportional to the  contribution to the static spin susceptibility from a band that dominantly couples to  $^{87}$Rb, we conclude that associated intrinsic susceptibility $\chi(T)$ is also significantly temperature dependent. This would be highly unusual for the Pauli susceptibility of the Fermi liquid in two- and three-dimensions. 
On the other hand, large and  temperature dependent $\chi(T)$ is not uncommon to 1D metals \cite{LiMoS_PRl}. Calculations for the Hubbard model of interacting electrons in 1D in fact predict a maximum in  $\chi(T)$ at $T\sim 0.18t$ ($t$ is the hopping matrix element) followed by an inflection at $T\sim 0.1t$ \cite{Nelisse1999}. While the comparison between the experimental and theoretical data can be at this stage only at a qualitative level, the temperature dependent $K_{87}$  implies the dominant coupling of $^{87}$Rb to the 1D bands and explains why $K_{87}$ does not follow bulk spin susceptibility (Fig.~S1 in \cite{supplemental}) with contributions from both 1D and 3D bands \cite{Yang2019}.

\begin{figure} [tb]
  \centering
\includegraphics[width=1\linewidth]{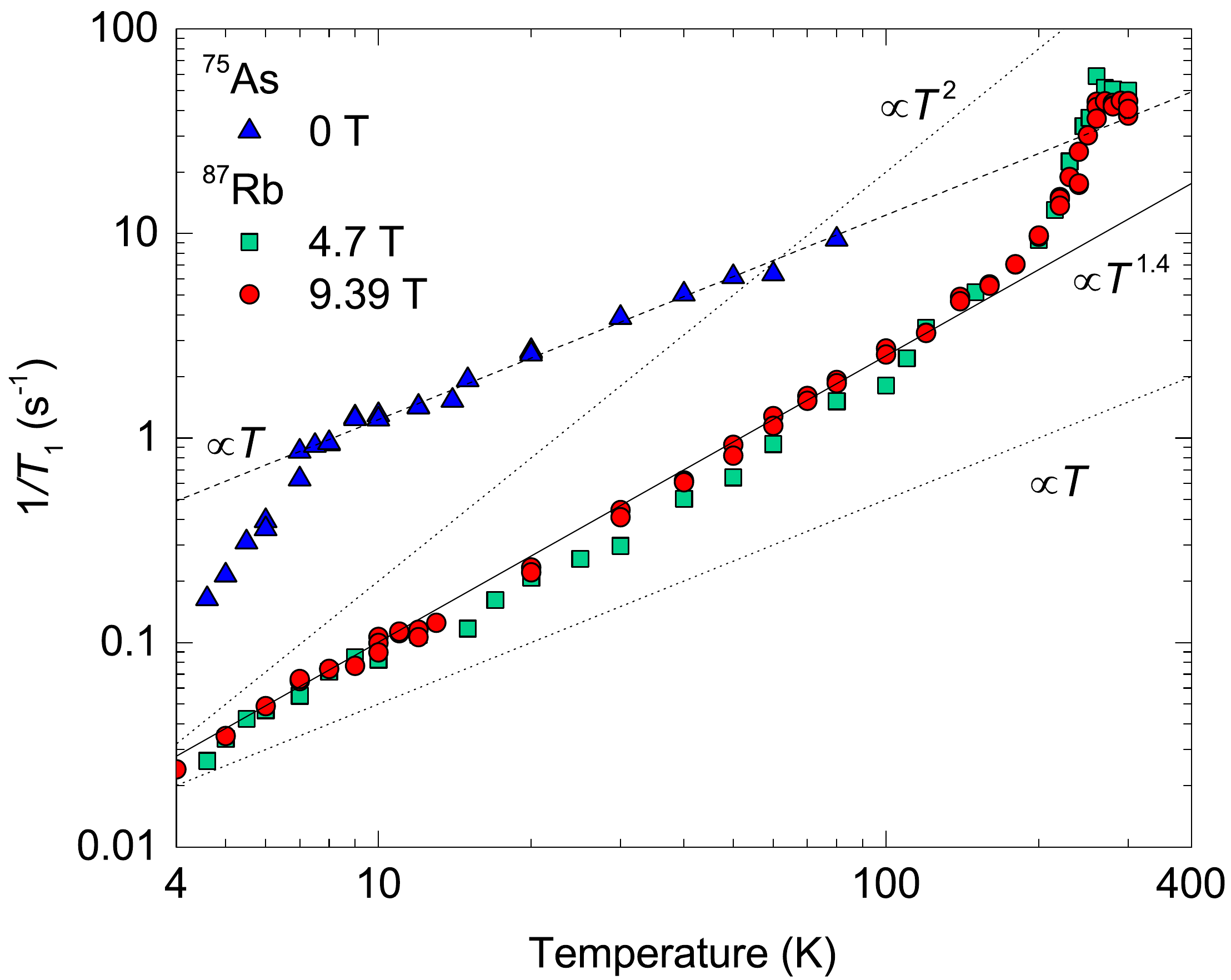}
  \caption{
Temperature dependence of  the zero-field $^{75}$As (blue triangles) and $^{87}$Rb $1/T_1$ measured in 4.7~T (green squares) and 9.39~T (red circles) for the Rb$_2$Mo$_3$As$_3$  powder. The power-law dependence $1/T_1\propto T^p$ with $p=1.4$ (solid line) for $^{87}$Rb data demonstrates the TLL physics of 1D bands. For comparison, linear and quadratic dependencies are denoted with dotted lines, respectively.  $^{75}$As nuclei, which are predominantly coupled to the 3D electronic band, show the conventional  $1/T_1\propto T$ behavior (dashed line).
  }
  \label{fig2}
\end{figure}

In order to quantitatively discuss the possible TLL physics, we switch to  the $^{87}$Rb  $1/T_1$ (Fig.~\ref{fig2}),  which is proportional to the sum of the imaginary part of the electron spin
susceptibility $\chi'' (q,\omega_{\rm L})$ over the wave vector $q$ and calculated at $\omega_{\rm L}=2\pi\nu_{\rm L}$: $1/T_1T\propto \sum_{q}\vert a_{\rm Rb}(q) \vert^2{\chi'' (q,\omega_{\rm L})\over \omega_{\rm L}}$. This makes $1/T_1$ an extremely sensitive probe of the low-energy dynamics. 
$1/T_1$ has a weakly  pronounced  field-dependent maximum at $T\sim 250$~K  (Fig.~S2 in \cite{supplemental}), which coincides with the minimum in $\rho (T)$ and indicates the freezing out of the Rb$^+$ ion dynamics, most probably of Rb at the $1c$ position where the alkali metal vacancies are primarily located \cite{Kvacancies}. However, once such dynamics is  completely frozen on the time-scale of  $\omega_{\rm L}$ below 200~K,  a clear power-law temperature dependence of $1/T_1 \propto T^p$ is observed in 4.7~T and 9.39~T over more than a decade in temperature between $\sim 10$~K and 200~K. This is a signal that the low-energy dynamics is described by the critical phenomena of collective modes and is thus a  hallmark of the TLL where the quantum critical regime is universal.  
$1/T_1T$ shows no contribution from the ferromagnetic spin fluctuations  similar to those recently reported for the superconducting A$_2$Cr$_3$As$_3$ \cite{CrAs_FM_SC}. 
Finally, $1/T_1$ data does not follow  the Fermi-liquid-type Korringa relaxation  $1/T_1T\propto {K_{87}}^2$ either (Fig. S3 in \cite{supplemental}), which implies that the 3D part of the Fermi surface has a negligible contribution to the  $^{87}$Rb $1/T_1$. 
The $^{87}$Rb $1/T_1$  is thus completely dominated by the two 1D bands, which are responsible for the TLL behavior. On the other hand,  zero-field  $^{75}$As $1/T_1$ nuclear quadrupole resonance (NQR) measurements (Figs. S4 and S5 in \cite{supplemental})  show $1/T_1\propto T$ (Fig.~\ref{fig2}) and suggest that $^{75}$As sitting in the Mo$_3$As$_3$ chains predominantly couple to the part of the Fermi surface with a 3D character. Combined   $^{87}$Rb and $^{75}$As $1/T_1$ thus corroborate multi-band physics \cite{Yang2019} and the interplay of 1D and 3D bands in the studied system.

The power-law fit of  $^{87}$Rb  $1/T_1$ yields $p\approx 1.4$  for both 4.7~T and 9.39~T measurements. This is dramatically different from $p\approx 0.75$, determined in K$_2$Cr$_3$As$_3$ \cite{Imai_PRL} or $p=0.34$ in single-wall carbon nanotubes \cite{Ihara_2010}  and immediately suggests that the 1D bands of Rb$_2$Mo$_3$As$_3$ are in a different TLL regime. In general, TLL is described by separate spin and charge collective modes, described by the corresponding TLL interaction parameters $K_{\rm s}$ and $K_{\rm c}$, respectively.   Assuming the spin invariance in low fields, which gives $K_{\rm s} = 1$, we focus only on the charge mode, which is affected by the interactions \cite{Dora-CNT-PRL}. For the  single-band TLL case, $1/T_1T\propto T^{\eta/2-2}$ is predicted, where  $\eta = 2K_{\rm c}+2$  holds the information about the TLL parameter for the $2k_{\rm F}$ spin fluctuations. From the experiment we  get $K_{\rm c} = p = 1.4$, which puts the studied  system into the regime with dominant attractive interactions. 
The more general treatment of the case of two  1D  bands  has been developed in the context of single-wall carbon nanotubes \cite{Dora-CNT-PRL}. The number of charge and spin excitation modes doubles 
and thus additional independent measurements are required to extract all the TLL parameters \cite{Ihara_2010}.  We stress though, that the simple one-band TLL limit is valid also for  two-band system, if the two corresponding charge modes are decoupled.
Regardless of these uncertainties,  
the finding that $K_{\rm c}>1$ is important because it demonstrates for Rb$_2$Mo$_3$As$_3$ that the effective interactions are attractive. This opens a path to superconductivity, whether with singlet or triplet symmetry even in the presence of weak disorder, depending on the strength of the scattering processes \cite{Anderson1D,  schulz1998, Ohta2008, pouget2012}.

\begin{figure} [t]
  \centering
\includegraphics[width=1.0\linewidth]{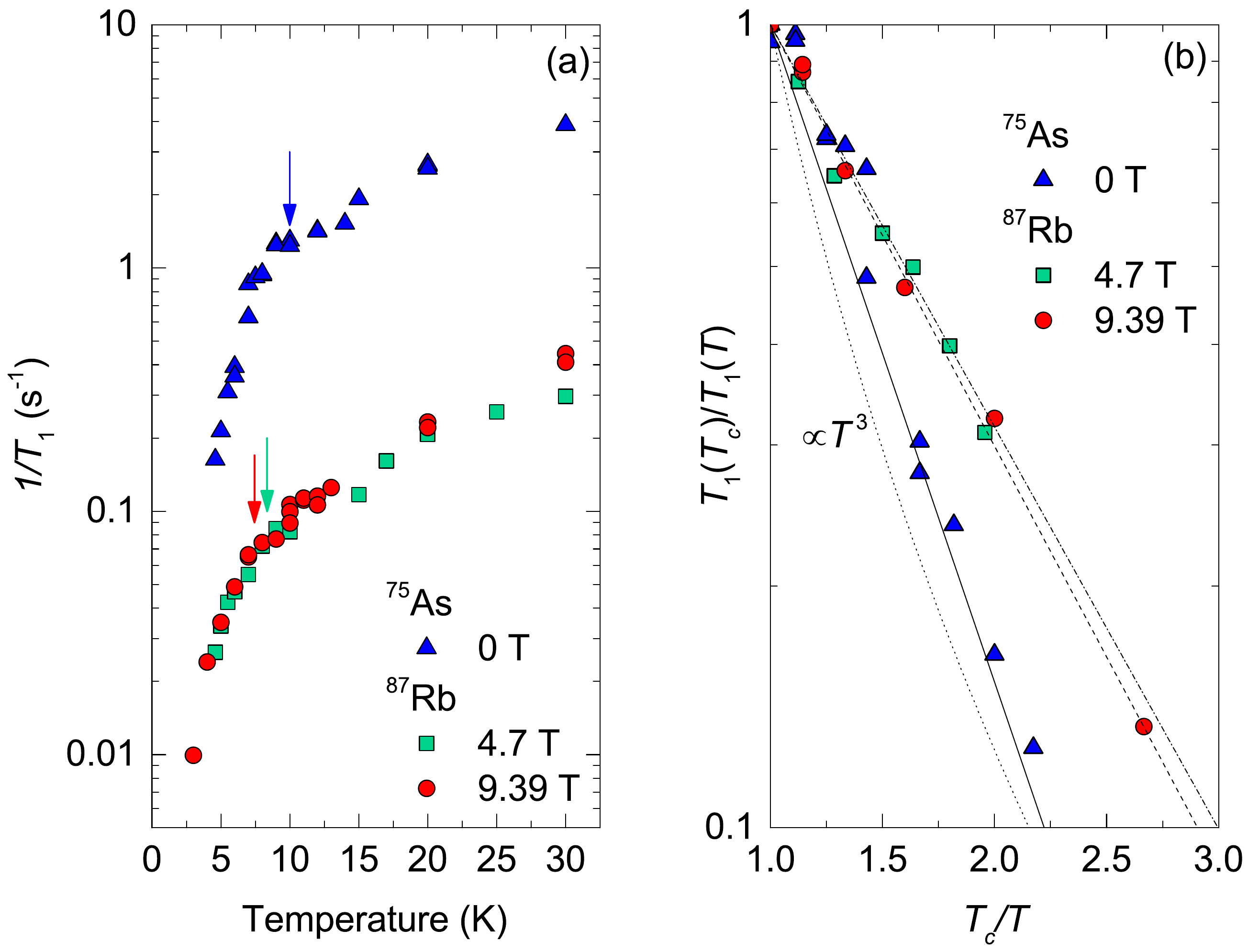}
  \caption{
  (a) Temperature dependence of $^{87}$Rb $1/T_1$ for Rb$_2$Mo$_3$As$_3$ measured in 4.7~T (green squares) and 9.39~T (red circles) close to $T_{\rm c}$.  Blue triangles stand for the zero-field $^{75}$As $1/T_1$ values. Vertical arrows mark the superconducting transition temperatures. 
  (b)  Normalized $^{87}$Rb relaxation rates, $T_1(T_{\rm c})/T_1(T)$, {\em vs} inverse temperature, $T_{\rm c}/T$,  in 4.7~T (green squares) and 9.39~T (red circles). For comparison, zero-field $^{75}$As $T_1(T_{\rm c})/T_1(T)$ is also shown (blue triangles).
Dashed lines are fits  of  $^{87}$Rb $1/T_1$ to an activated temperature dependence yielding $2\Delta_0/k_{\rm B}T_{\rm c}\approx 2.4$, while solid line is a fit of $^{75}$As $1/T_1$ to the same model with $2\Delta_0/k_{\rm B}T_{\rm c}\approx 3.7$ (Table SI in \cite{supplemental}).
$1/T_1\propto T^3$ dependence 
is shown by a dotted line.
  }
  \label{fig3}
\end{figure}

The onset of superconductivity below $T_{\rm c}\approx 8$~K   is  barely seen in the line shift  of   the $^{87}$Rb NMR spectra measured in 9.39~T (Fig.~\ref{fig1}c), but can be  deduced from the wipeout effect (Fig.~\ref{fig1}d).  In the temperature dependence of  $^{87}$Rb $1/T_1$, the onset of superconductivity is seen as a sudden suppression of $1/T_1$ 
(Fig.~\ref{fig3}a) without any Hebel-Slichter coherence peak, which would be a hallmark of the conventional  Bardeen–Cooper–Schrieffer (BCS) s-wave superconductivity, but is common to unconventional superconductors, including  related iron-pnictide superconductors \cite{Ma_2013} and  A$_2$Cr$_3$As$_3$ \cite{Imai_PRL, CrAs_FM_SC}. 
In Rb$_2$Mo$_3$As$_3$, $1/T_1$ apparently follows an activated temperature dependence, $1/T_1\propto \exp(-\Delta_0/k_{\rm B}T)$ down to $T_{\rm c}/T=2.7$ (Fig.~\ref{fig3}b).
Attempts to fit $1/T_1$ to the power law  $T^{n}$ for $T_{\rm c}>T>T_{\rm c}/2.7$ with $n\sim 3-5$ result in a  worse fit. 
We thus employ $1/T_1\propto \exp(-\Delta_0/k_{\rm B}T)$ to estimate the superconducting gap $\Delta_0 \approx 0.9$~meV (Fig.~\ref{fig3}b and Table~SI in \cite{supplemental}).  Using $T_{\rm c}\approx 8$~K  deduced from the onset of superconductivity in $\rho (T)$ (Fig.~\ref{fig0}d), we can then calculate the energy gap-$T_{\rm c}$ ratio $2\Delta_0/k_{\rm B}T_{\rm c}=2.4$, well below the standard BCS value of $2\Delta_0/k_{\rm B}T_{\rm c}\geq 3.52$. Equivalent analysis and conclusions hold also for measurements in $4.7$~T.
Remarkably, the temperature dependence of the zero-field $1/T_1$ for the $^{75}$As, which couples to the 3D part of the Fermi surface, can still be described by the thermally activated dependence without Hebel-Slichter coherence peak, but with a larger $\Delta_0 = 1.6$~meV and $2\Delta_0/k_{\rm B}T_{\rm c} = 3.7$.

Complementary $^{87}$Rb NMR and $^{75}$As NQR data reveal that Rb$_2$Mo$_3$As$_3$ has to be treated as a multiband system comprising 1D and 3D bands. It is thus reasonable to assume a two-gap scenario for the superconductivity.  
However, such a two-gap scenario could not easily explain the difference between the extracted $2\Delta_0/k_{\rm B}T_{\rm c}$ from the $^{75}$As and $^{87}$Rb data and {\red could be seen if the gap averaging due to impurity scattering is weak \cite{TwoGap}.} As the resistivity data (Fig. \ref{fig0}c) indicate the presence of disorder, such an interpretation of the NMR and NQR data is {\red less likely} to  apply to  Rb$_2$Mo$_3$As$_3$.
On the other hand, the smallness of the extracted $\Delta_0$  for the $^{87}$Rb NMR  may result from the vortex core relaxation \cite{Rigamonti}. In this case, the measured $^{87}$Rb NMR $1/T_1$ is a weighted average of the relaxation in the vortex core and in the inter-vortex region. If that is the case, then we find the marginal difference between the 4.7~T and 9.39~T data  fairly surprising. 

The absence of a noticeable shift in $^{87}$Rb NMR upon cooling through the superconducting transition (Fig. \ref{fig1}c) may imply the elusive  spin-triplet superconductivity~\cite{Lee_triplet_SC}, which is compatible with the theoretical prediction of the triplet  superconducting fluctuations \cite{giamarchi2003, Anderson1D,  schulz1998, Ohta2008, pouget2012} in the TLL regime of attractive interactions. 
Experimentally, the superconductivity should persist at fields exceeding the paramagnetic limit, as it was indeed previously observed in A$_2$Cr$_3$As$_3$ samples \cite{Cao_PRX2015, Tang_2017}. However, in triplet superconductors, the   power-law scaling of $1/T_1$ with $n\sim 3-5$ is expected~\cite{SW-triplet}, which has yet to be tested for Rb$_2$Mo$_3$As$_3$ with measurements to even lower temperatures. 
Finally, the extracted values of $2\Delta_0/k_{\rm B}T_{\rm c}$ may be affected also by the quenched disorder perturbing TLL near the superconductor-to-insulator transition leading to situations when the length scale of the inhomogeneity is comparable to or larger than the coherence length \cite{Disorder-Low_gap, PRB_disorder}.   In Rb$_2$Mo$_3$As$_3$, the source of disorder could be connected to the Rb$^+$ $1c$ sites, whereas the effective attractive interactions within the TLL place this system very close to the metal(superconductor)-to-insulator critical point. The increase in $\rho(T)$ with decreasing temperature below $T_{\rm min}$  (Fig.~\ref{fig0}c) seems to corroborate this suggestion, but more experiments {\red to even lower temperatures} are needed to fully understand the relation between the disorder,  the size and the sign of interactions defining the  TLL and the emerging bulk superconductivity in the A$_2$Mo$_3$As$_3$ family.

In conclusion, we have studied Rb$_2$Mo$_3$As$_3$ composed of the assembly of weakly coupled Mo$_3$As$_3$ chains and have shown that the  1D bands adopt TLL behavior over a strikingly broad temperature range. 
A power-law dependence of $^{87}$Rb $1/T_1$ demonstrates the realization of the TLL with $K_{\rm c}>1$ which has been rarely accessible before in realistic physical systems.  The multiband electronic structure  probed by $^{75}$As and $^{87}$Rb nuclei, the vortex core relaxation and the underlying TLL characterized by the effective attractive  coupling are discussed  in connection to the unusual superconductivity marked by the  temperature independent Knight shift, the absence of the Hebel-Slichter coherence peak and the reduced superconducting gap.

{\em Acknowledgements} D.A. acknowledges the financial support of the Slovenian Research Agency through J1-9145 and P1-0125 grants.
 The work at UT Dallas is supported  by US Air Force Office of Scientific Research (AFOSR) No. FA9550-15-1-0236 and FA9550-19-1-0037.

%



\end{document}



\title{Superconductivity in the regime of attractive interactions in the Tomonaga-Luttinger liquid
\\
Supplemental Material}

\author{\v Z. Gosar}
\affiliation{Institute Jo\v zef Stefan, 
Jamova c.~39, 1000 Ljubljana, Slovenia}
\affiliation{Faculty of mathematics and physics, University of Ljubljana, 
Jadranska c.~19, 1000 Ljubljana, Slovenia}
\author{N. Jan\v sa}
\affiliation{Institute Jo\v zef Stefan, 
Jamova c.~39, 1000 Ljubljana, Slovenia}
\author{T. Arh}
\affiliation{Institute Jo\v zef Stefan, 
Jamova c.~39, 1000 Ljubljana, Slovenia}
\author{P. Jegli\v c}
\affiliation{Institute Jo\v zef Stefan, 
Jamova c.~39, 1000 Ljubljana, Slovenia}
\author{M. Klanj\v sek}
\affiliation{Institute Jo\v zef Stefan, 
Jamova c.~39, 1000 Ljubljana, Slovenia}
\author{H. F. Zhai}
\affiliation{The University of Texas at Dallas,  800 West Campbell Road Richardson, Texas 75080-3021, USA}
\author{B. Lv}
\affiliation{The University of Texas at Dallas,  800 West Campbell Road Richardson, Texas 75080-3021, USA}
\author{D. Ar\v con}
\email{denis.arcon@ijs.si}
\affiliation{Institute Jo\v zef Stefan, 
Jamova c.~39, 1000 Ljubljana, Slovenia}
\affiliation{Faculty of mathematics and physics, University of Ljubljana, 
Jadranska c.~19, 1000 Ljubljana, Slovenia}

\maketitle


\begin{table}[h!]
\begin{center}
\begin{tabular}{ c|c|c|c|c }
Nuclei & Magnetic field (T) & $T_{\rm c}$ (K)  & $\Delta_0$ (meV) & $2\Delta_0/k_{\rm B}T_{\rm c}$  \\
\hline
$^{75}$As & 0 & 10 & 1.62 & 3.7\\
$^{87}$Rb & 4.7 & 9 & 0.89 & 2.4\\
$^{87}$Rb & 9.39 & 8 & 0.83 & 2.4\\
\end{tabular}
\end{center} 
\caption{Summary of fitting parameters for the spin-lattice relaxation rates in the superconducting state of Rb$_2$Mo$_3$As$_3$. The critical temperature $T_{\rm c}$ has been taken as an onset of the superconducting transition in the resistivity measurements (Fig. 1d in the main text), while the superconducting gap $\Delta_0$ has been determined from the fits of the spin-lattice relaxation rates to the thermally activated behavior (Fig. 4b in the main text). }
\label{table1}
\end{table}

\begin{figure} [h]
  \centering
 \includegraphics[width=0.90\linewidth]{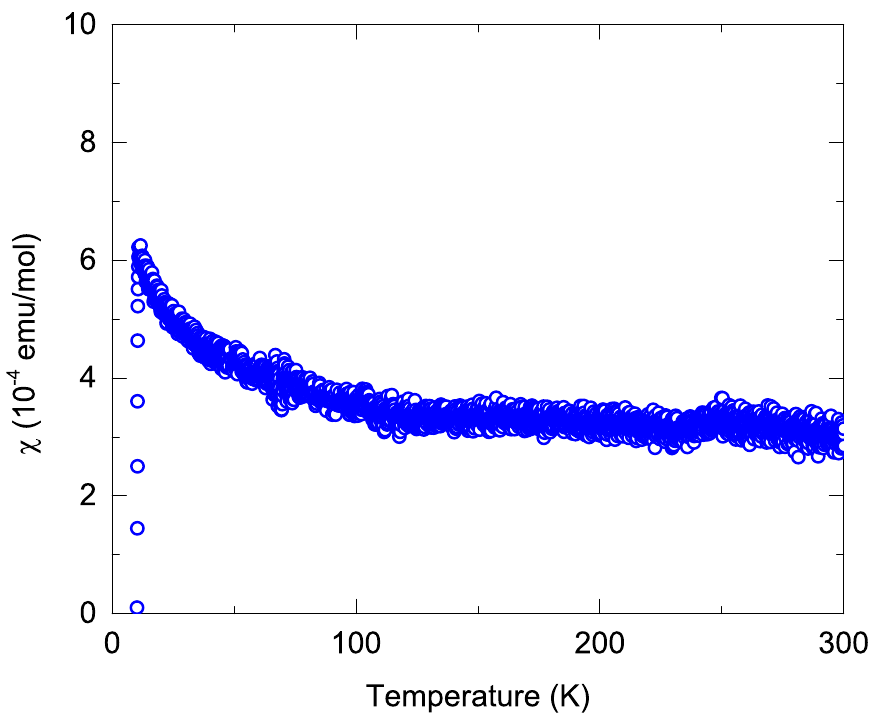}
  \caption{
Molar susceptibility of Rb$_2$Mo$_3$As$_3$ powder sample measured with the zero-field-cooling protocol and in the $0.1$~T magnetic field. Quantum Design Physical Property Measurement System (PPMS) magnetometer has been used in these experiments.} 
  \label{fig0}
\end{figure}

\begin{figure} [h]
  \centering
 \includegraphics[width=1.0\linewidth]{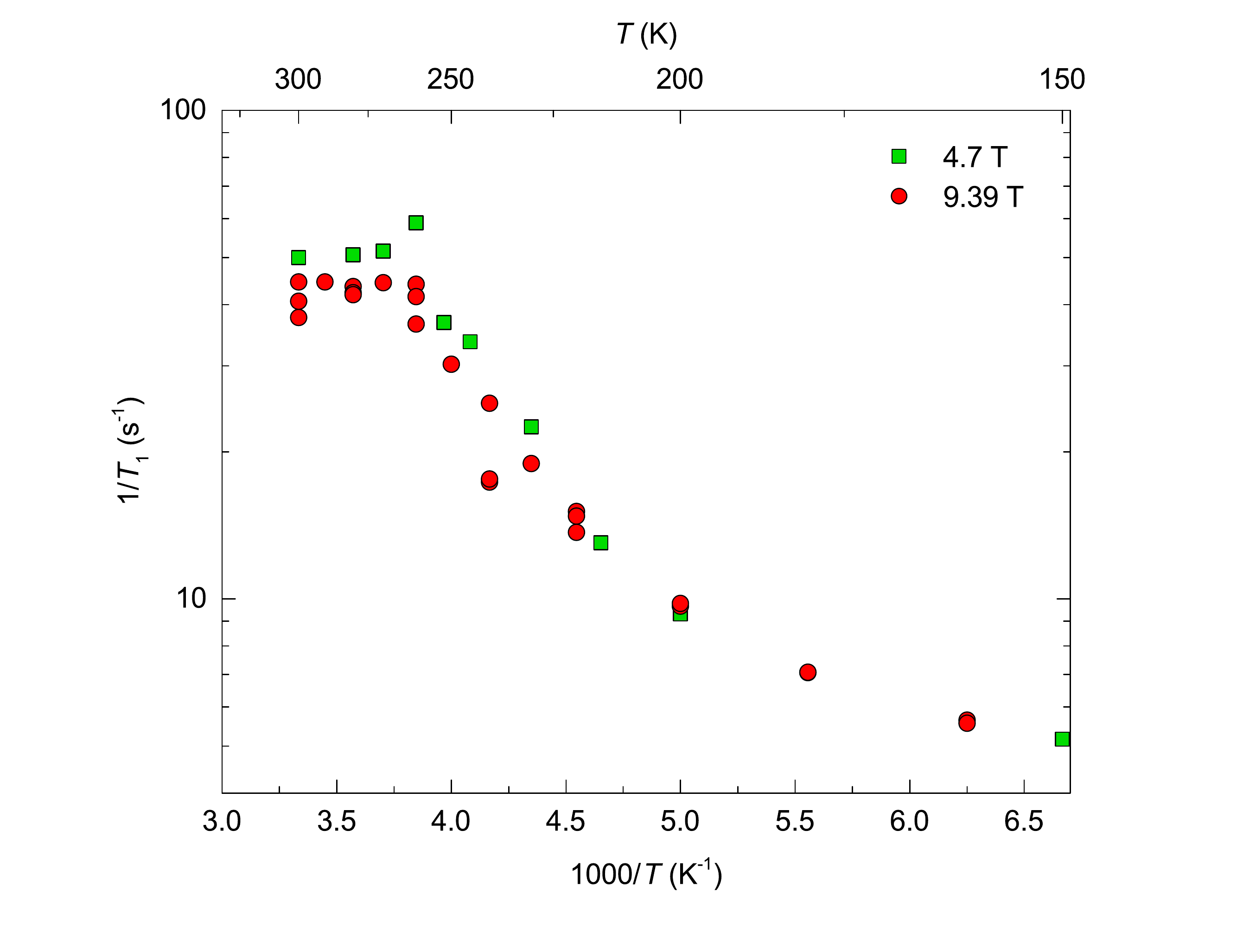}
  \caption{
$^{87}$Rb $1/T_1$ measured in Rb$_2$Mo$_3$As$_3$ powder sample shows enhancement in a lower magnetic field around the maximum, which is expected for the Bloembergen-Purcell-Pound-type relaxation.  
} 
  \label{figS2}
\end{figure}

\begin{figure} [h]

	  \centering
	  \includegraphics[width=\linewidth]{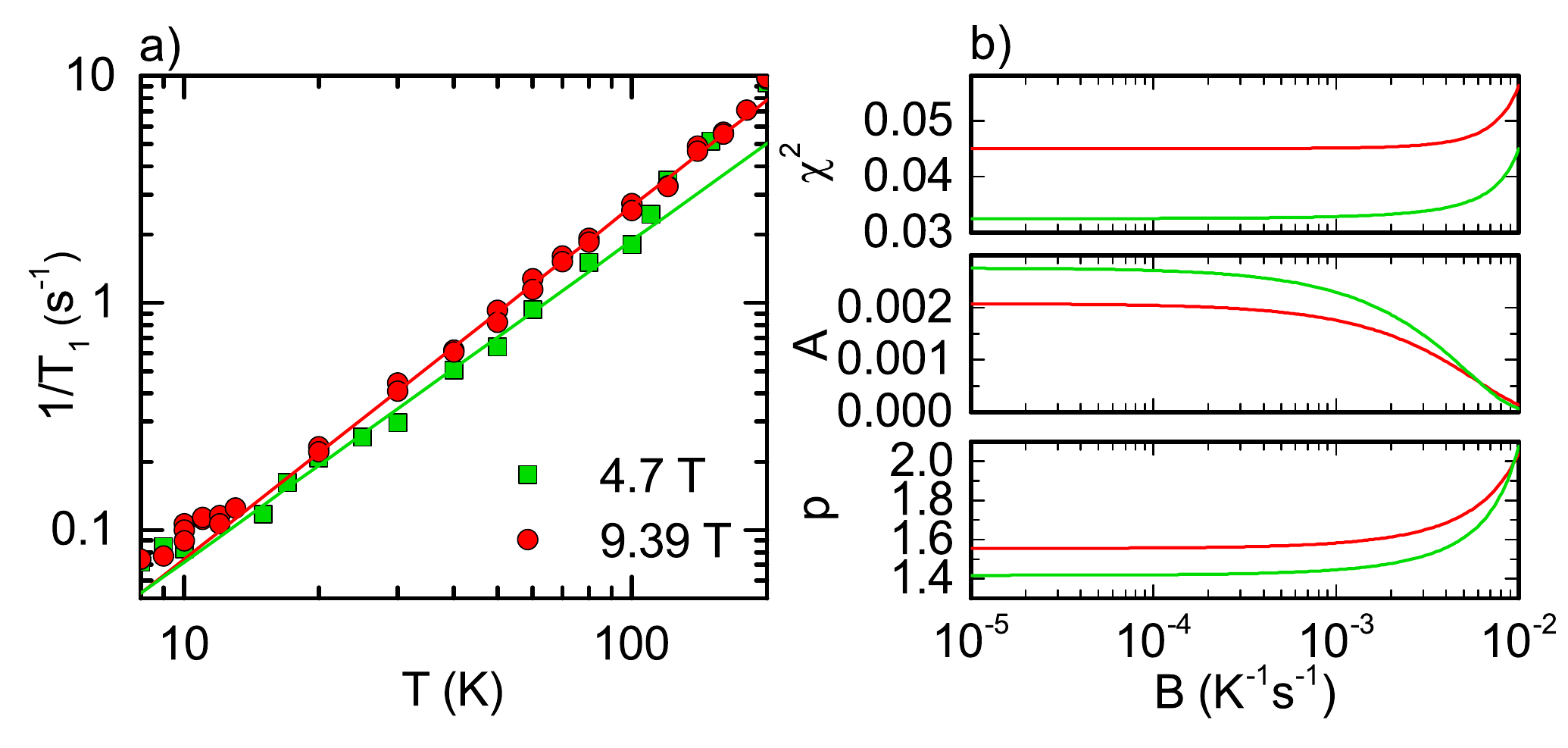}
  \caption{a) $^{87}$Rb spin-lattice relaxation rate $1/T_1$ measured in magnetic fields of $4.7\ \mathrm{T}$ (green squares) and $9.39\ \mathrm{T}$ (red circles). The data was fitted to a model $1/T_1 = AT^p + BT$ in the range $20\ \mathrm{K}$~--~$100\ \mathrm{K}$ where we safely avoided possible non-universal regime (solid lines). The first term includes 1D TLL power-law dependence while the second stands for the 3D Korringa-type relaxation. With the only constraint $B > 0$, $B$ converges to $0$ in both fields ($B = 4.12 \times 10^{-18}\ \mathrm{K^{-1}s^{-1}}$ and $B = 4.68 \times 10^{-14}\ \mathrm{K^{-1}s^{-1}}$ respectively), meaning the 3D contribution is negligible. The corresponding $A$ and $p$ parameters are $A = 2.76\times 10^{-3}$, $p = 1.42$ and $A = 2.07\times 10^{-3}$, $p = 1.55$, respectively.
  b) $\chi^2$ value  of the fit (top), $A$ and $p$ parameters of the fit (middle and bottom) to  $1/T_1 = AT^p + BT$ with different fixed $B$ values.  When $B$ term becomes relevant, for $B  > 10^{-3}\ s^{-1}K^{-1}$, $\chi^2$ rapidly worsens. }
  \label{figS5}
\end{figure}

\begin{figure} [h]
  {\centering
 \includegraphics[width=1.0\linewidth]{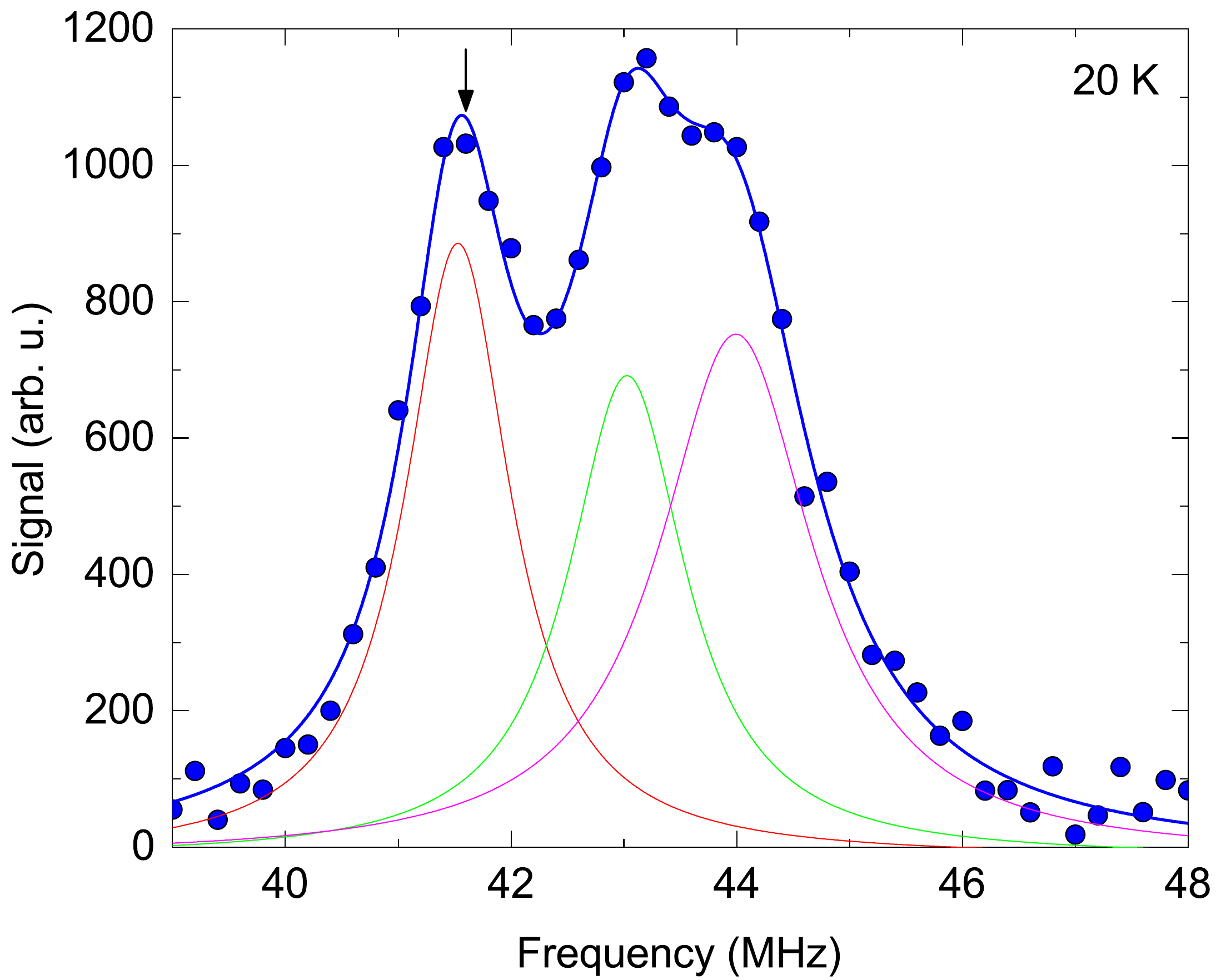}
 }
  \caption{
$^{75}$As NQR spectrum of Rb$_2$Mo$_3$As$_3$ powder measured at $T=20$~K. Thin red, green and magenta lines are individual  Lorentzian components  used to deconvolute $^{75}$As NQR spectra.  Solid thick blue line is their sum. Arrow indicates the frequency at which spin-lattice relaxation rate measurements were conducted. 
} 
  \label{figS3}
\end{figure}

\begin{figure} [h]
  \begin{center}
 \includegraphics[width=1.0\linewidth]{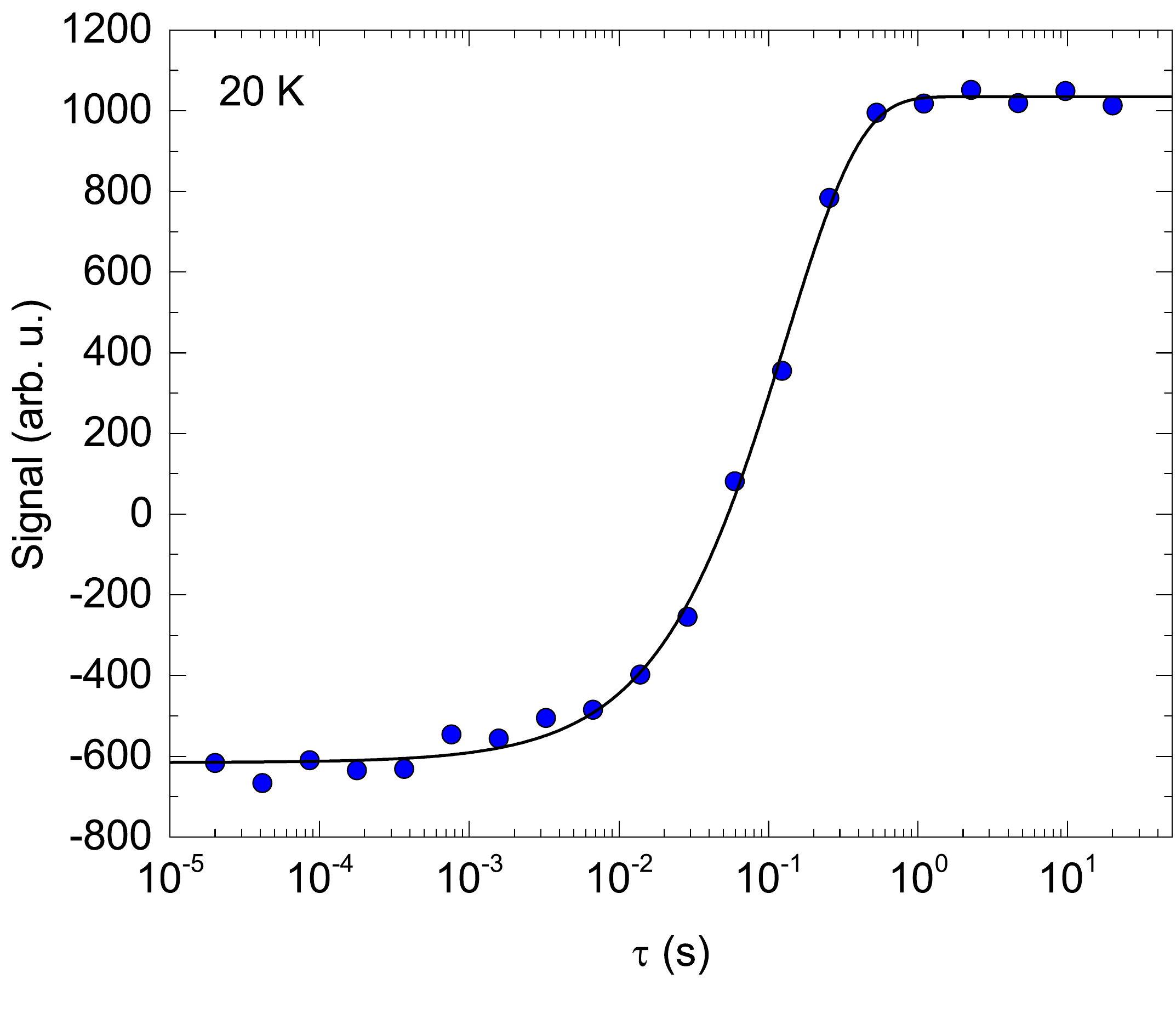}
   \end{center}
  \caption{
The $^{75}$As NQR signal intensity as a function of delay time $\tau$ in the spin-lattice relaxation rate measurements. Standard inversion-recovery pulse sequence $\pi - \tau - \pi /2 - \tau_1 -\pi - \tau_1 - {\rm echo}$ (pulse lengths were typically $\pi /2 = 3.1$~$\mu$s and the delay time $\tau_1 = 100$~$\mu$s) was used. The $^{75}$As magnetization recovery curve was fitted to the expression  for $^{75}$As $I=3/2$: $M_z(\tau) = M_0 \left( 1-(1+s)\exp [-(3\tau /T_1)^\alpha] \right)$. Here $M_0$ is the signal amplitude, $s$ measures the efficiency of the nuclear magnetization inversion, and $\alpha \approx 0.9$ is the stretching exponent.   
} 
  \label{figS4}
\end{figure}

